\documentclass[aps,prb,twocolumn,amsmath,amssymb]{revtex4}

\usepackage{graphicx}
\usepackage{dcolumn}
\usepackage{bm}
\usepackage[caption=false]{subfig}
\usepackage{amssymb}
\usepackage{amsmath}
\usepackage{verbatim}
\usepackage{mathtools}
\usepackage{amsfonts}
\usepackage{physics}
\usepackage{siunitx}

\usepackage{hyperref}
\usepackage{color}

\def\beq{\begin{equation}}
\def\eeq{\end{equation}}

\begin{document}

\title{Ce and Dy substitutions in Nd$_{2}$Fe$_{14}$B: site-specific magnetic anisotropy from first-principles}
	
	\author{James Boust$^{1}$, Alex Aubert$^{2}$, Bahar Fayyazi$^{2}$,  Konstantin P. Skokov$^{2}$,  Yurii Skourski$^{3}$, Oliver Gutfleisch$^{2}$ and Leonid V. Pourovskii$^{1,4}$}

	\affiliation{
		$^1$CPHT, Ecole Polytechnique, CNRS,
		Institut polytechnique de Paris, 91128 Palaiseau, France \\
		$^2$Functional Materials, TU Darmstadt, 64287 Darmstadt, Germany\\
		$^3$Dresden High Magnetic Field Laboratory (HLD-EMFL), Helmholtz-Zentrum Dresden Rossendorf, D-01328 Dresden, Germany\\
		$^4$Coll\`ege de France, 11 place Marcelin Berthelot, 75005 Paris, France}
        
\begin{abstract}
A first-principles approach combining density functional and dynamical mean-field theories  in conjunction with a quasi-atomic
approximation for 
the strongly localized 4$f$ shell is applied to Nd$_{2}$Fe$_{14}$B-based hard magnets in order to evaluate  crystal-field and exchange-field parameters at rare-earth sites and their corresponding single-ion contribution 
to the magnetic anisotropy.
In pure Nd$_2$Fe$_{14}$B, our calculations reproduce the easy-cone to easy axis transition; theoretical  magnetization curves  agree quantitatively with experiment. Our study reveals that the rare-earth single-ion anisotropy in the "2-14-1" structure is strongly site-dependent, with the $g$ rare-earth site exhibiting a larger value. In particular, we predict that increased $f$ and $g$-site occupancy of $R=$ Ce and Dy, respectively, 
leads to an increase
of the magnetic anisotropy  of the corresponding (Nd,$R$)$_{2}$Fe$_{14}$B substituted compounds. 

\end{abstract}
\maketitle

\section{Introduction}

High-performance permanent magnets are key components of numerous energy-efficient technologies which have today to meet an increasing need, such as wind generators and electrical motors\cite{Skokov2018,Coey2011,Hono2018,Skokov2018bis}. Understanding and optimizing their intrinsic properties arising at the atomic level as well as their extrinsic properties due to the microstructure, is therefore crucial economically- and environmentally-wise\cite{Skokov2018}. The market is dominated by rare-earth (R) transition-metal (M) intermetallics\cite{Skokov2018,Coey2020} whose intrinsic properties are generally understood within the two sub-lattice picture\cite{Kuzmin2007}: the itinerant 3$d$ electrons of a late transition metal, such as Fe and Co, provide the compound with strong permanent magnetism behavior (i.e. large magnetization and Curie temperature) while the magnetic hardness mainly stems from the R localized 4$f$ shells, whose strong spin-orbit coupling converts the crystal field anisotropy into a magnetic one. The interplay between these two sub-lattices is governed by the M induced exchange field, the 4$f-$4$f$ interaction usually being much weaker\cite{Kuzmin2004,Fuerst1986,Nicklow1976} and neglected. 

The most widely used high-performance permanent magnet in the industry is the  Nd$_{2}$Fe$_{14}$B (2-14-1) intermetallic, in which Nd is often partially substituted by heavy Rs such as Dy or Tb to further enhance the magnetic properties\cite{Skokov2018,Coey2020,Hono2018,Skokov2018bis}. However, substantial efforts are being made to find alternative hard magnets with reduced critical R content  because of their economical as well as environmental costs\cite{Coey2011,Hono2018,Skokov2018bis}. For instance, there is growing interest in  Ce, which is far more abundant than Nd\cite{Skokov2018bis,Chouhan2018,liao2022}.

Within this context, understanding the intrinsic properties of complex 2-14-1 systems with partial Nd substitution would help optimizing the design of new hard magnets. It requires, 
however, a quantitative description of  crystal-field effects, which induce the single-site rare-earth magnetic anisotropy and, thus, the magnetic hardness in these systems. A reliable evaluation of Crystal Field Parameters (CFPs) is  a difficult task both from the experimental and theoretical perspectives. Indeed, extracting them by fitting experimental high field magnetization curves\cite{Yamada1988,Cadogan1988,Kostyuchenko2020} has a limited predictive power, as it usually neglects some CFPs, and requires the use of single crystals. Furthermore, care must be taken to properly separate contributions from the M and R sub-lattices\cite{Ito2016}. An alternative route is to compute these CFPs from first principles which is also a notorious challenge, notably due to the localized and strongly correlated nature of R 4$f$ states which standard Density Functional Theory (DFT) fails to describe properly. Nd$_{2}$Fe$_{14}$B-based systems are especially hard to treat due to a large unit cell with two inequivalent R sites. First works towards \textit{ab initio} methods were still partially relying on the crude point-charge electric model\cite{Zhong1989,Zhong1989bis}. Several DFT-based approaches have been so far developed and applied to various R-M intermetallics \cite{Novak1994,Hummler1996,Novak1996,Divis2005,Staunton2019}, but they are usually relying on open-core-like treatment of R 4$f$ shells neglecting R-M hybridization. The latter can have a significant impact on CFPs and magnetic properties\cite{Pourovskii2020}. Regarding 2-14-1 systems, after early studies focusing on the '20' CFP\cite{Tanaka2011,Tanaka2011_2}, authors of Refs. \onlinecite{Yoshioka2015,Tsuchiura2018,Yoshioka2018,Sato2021} used a Wannier function based approach to compute all CFPs in various compounds but hybridization was taken into account in an approximate way\cite{Yoshioka2015}.

Recently, Delange \textit{et al}.\cite{Delange2017} introduced a new approach to the first-principles calculation of CFPs in such intermetallics by treating the 4$f$ shell, in the framework of the Dynamical Mean-Field Theory (DMFT)\cite{Metzner1989,Georges1992}, within the  quasi-atomic Hubbard-I (HI) approximation\cite{Hubbard1963}. This DFT+DMFT\cite{Aichhorn2009} method removes the unphysical 4$f$ contribution to CF by an averaging scheme. It was successfully applied to RCo$_{5}$ compounds by Pourovskii \textit{et al}. \cite{Pourovskii2020} who explained the 40-year-old mystery of the measured zero-temperature Nd magnetic moment in NdCo$_{5}$\cite{Alameda1982} being frozen below saturation by a large high-rank '66' CFP arising from the hybridization with neighboring Co atoms. Hybridization can be implicitly taken into account in this methodology by a proper choice of the 4$f$ Wannier projection window.

Building on this success, we apply in this work the same methodology to 2-14-1 systems. Comparison of our computed CFPs and resulting magnetic properties with experimental data shows that this essentially \textit{ab initio} approach can accurately capture intrinsic magnetic properties of the parent Nd$_{2}$Fe$_{14}$B compound such as the temperature evolution of the spontaneous magnetization direction and magnetization curves. We then focus on site-specific partial Nd substitutions by either Dy or Ce. Our main results are: (i) the CFPs on a given inequivalent R site 
are not sensitive to substitutions on the second 
R site  and (ii) the R single-ion anisotropy (SIA) is strongly site dependent. We therefore predict that optimization of site occupancy in Ce and Dy substituted Nd$_{2}$Fe$_{14}$B could lead to an increase, though moderate, of magnetic anisotropy. We also
measure the impact of Ce substitution on magnetocrystalline anistropy energy (MAE) on single crystals and compare it with theoretical predictions.

The paper is organized as follows : in Section II, we review the theoretical methodology and experimental procedures used in this work; in Section III, we present our results on pure Nd$_{2}$Fe$_{14}$B as well as Ce and Dy substituted compounds.

\section{Theoretical approach}
\subsection{Electronic structure calculations}\label{subsec:estruct}

The electronic structure calculations of rare-earth transition metal intermetallics are carried out using the DFT+HI approach of Refs. \onlinecite{Delange2017} and \onlinecite{Pourovskii2020}. Within this self-consistent over the charge density DFT+DMFT scheme\cite{Aichhorn2009}, the local-spin density approximation (LSDA) is employed to describe the M magnetism while the quasi-atomic HI\cite{Hubbard1963} approximation for the DMFT\cite{Metzner1989,Georges1992} impurity problem is used to describe the R 4$f$ shell. La is treated within LSDA as its 4$f$ shell is empty (La$^{3+}$). Spin-orbit coupling is included within the standard second-variation procedure as implemented in Ref. \onlinecite{Wien2k}. The experimental crystal structure (space group P4$_{2}$/mnm) and lattice parameters are used\cite{Herbst1991} throughout; lattice distortion with temperature\cite{Givord1985} are neglected. In order to evaluate the lattice parameters of the mixed ($RR'$)$_2$Fe$_{14}$B systems from those of the corresponding pure compounds we employ Vegard's law. Calculations are performed with the M magnetic moment aligned along the [001] direction. The approach is implemented within the full-potential linearized augmented planewave
(FP-LAPW) band structure code Wien2k\cite{Wien2k,Wienk_2} in conjunction with the TRIQS library\cite{triqs_main,triqs_dft_tools}.

R 4$f$ Wannier orbitals are constructed from the Kohn-Sham (KS) bands enclosed in an energy window of size $4$~eV. As in Ref. \onlinecite{Pourovskii2020}, for Nd, whose KS 4$f$ states are pinned at the Fermi level $E_F$, this window is centered around $E_F$; for Dy, as the 4$f$ KS bands move towards lower energies during the self-consistent calculation, the window is centered around the central weight of the 4$f$ partial density of states. 
This window should be large enough to enclose most of the 4$f$ KS states but small enough to capture the hybridization with M atoms which, as shown in Ref. \onlinecite{Pourovskii2020}, is crucial to the proper evaluation of CFPs.

Within the HI approximation, the DMFT impurity problem is reduced\cite{Lichtenstein1998} to the diagonalization of the following quasi-atomic Hamiltonian for the 4$f$ shell:
\begin{equation}
    \hat{H}_{at}=\hat{H}_{1el}+\hat{H}_U=\sum_{mm'\sigma\sigma'}\epsilon_{mm'}^{\sigma\sigma'}\hat{f}^{\dagger}_{m\sigma}\hat{f}_{m'\sigma'}+\hat{H}_{U}.\label{Hat}
\end{equation}
Here, $\hat{f}_{m'\sigma'}$ is the annihilation operator for the 4$f$ orbital $m'\sigma'$, $\hat{H}_{U}$ is the Coulomb repulsion Hamiltonian and $\epsilon_{mm'}^{\sigma\sigma'}$ is the one-electron level-position matrix:
\begin{equation}
    \epsilon=-(\mu+\Sigma_{DC})\mathbb{I}+\sum_{\bold{k}\in BZ}P_{\bold{k}}H_{KS}^{\bold{k}}P_{\bold{k}}^{\dagger}
\end{equation}
where $\mathbb{I}$ is the identity matrix, $\mu$ is the chemical potential, $\Sigma_{DC}$ is the double-counting correction computed here within the Fully-Localized Limit (FLL) with nominal atomic occupancy\cite{Pourovskii2007}, $H_{KS}^{\bold{k}}$ is the KS Hamiltonian matrix and $P_{\bold{k}}$ is the projection matrix between the 4$f$ Wannier and KS spaces\cite{Aichhorn2009}.

The Coulomb interaction $\hat{H}_{U}$ is specified by the parameters $J_{H}=0.77$ ($0.99$) eV (as measured by optical spectroscopy\cite{Carnall1989}) and $U=6$ ($7$)~eV for Nd (Dy) respectively.

\subsection{Single-ion model for the 4$f$ Hamiltonian}\label{subsec:SIM}

Once the DFT+HI scheme described in the previous sub-section has converged, the CFPs, spin-orbit coupling and exchange field are extracted by fitting the converged one-electron part $\hat{H}_{1el}$ of the quasi-atomic Hamiltonian (\ref{Hat}) onto the form expected within the single-ion model (which neglects 4$f$-4$f$ interactions). Namely, for a R ion embedded in a 2-14-1 crystal with ferromagnetically aligned M moments, it reads: 
\begin{equation}
    \hat{H}_{1el}=E_{0}\hat{\mathbb{I}}+\lambda\sum_{i}\hat{s}_{i}\hat{l}_{i}+\hat{H}_{ex}+\hat{H}_{CF}+\hat{H}_{ext}.
    \label{H4f}
\end{equation}
$E_{0}$ is an energy shift, $\lambda$ is the spin-orbit coupling, $\hat{H}_{ex}=2\mu_{B}\bold{B}_{ex}(T)\cdot \bold{\hat{S}}_{4f}$ is the M-R exchange interaction with $\bold{B}_{ex}(T)$ being the exchange field at temperature $T$ and $\bold{\hat{S}}_{4f}$ the 4$f$ spin operator. The M-R exchange interaction is a multi-orbital coupling as it has been shown to be mediated by the moment of R5$d$6$s$ orbitals\cite{Herbst1991,Coey1996}. The exchange field $\bold{B}_{ex}(T)$ used in the single-ion model captures the combined effects of the inter-atomic M3$d-$R5$d$6$s$ and intra-atomic R5$d$6$s-$R4$f$\cite{Frietsch2015,Pivetta2020} interactions as an effective mean-field acting on the R4$f$ shell.

$\hat{H}_{CF}$ is the crystal-field Hamiltonian, which, in 2-14-1 compounds, in the coordinate system x$\parallel$[100] and z$\parallel$[001], can be written as\cite{Yamada1988}:
\begin{align*}
    \hat{H}_{CF}  =L_{2}^{0} \hat{T}_{2}^{0}&+L_{2}^{-2} \hat{T}_{2}^{-2}\\
    & +L_{4}^{0} \hat{T}_{4}^{0}+L_{4}^{-2} \hat{T}_{4}^{-2}+L_{4}^{4} \hat{T}_{4}^{4}\\
     &+L_{6}^{0} \hat{T}_{6}^{0}+L_{6}^{-2} \hat{T}_{6}^{-2}+ L_{6}^{4}\hat{T}_{6}^{4}+L_{6}^{-6}\hat{T}_{6}^{-6}
\end{align*}
where $\hat{T}_{k}^{q}$ are the Hermitian combination of Wybourne's operators\cite{Wybourne1965} with the same notations as Ref. \onlinecite{Delange2017} and $L_{k}^{q}$ are the CFPs. As often done in the literature, we will use the Stevens convention\cite{Stevens1952} of the CFPs throughout, $A_{k}^{q}\langle r^{k}\rangle=\lambda_{kq}L_{k}^{q}$ ($\lambda_{kq}$ are tabulated in Ref. \onlinecite{Mulak2000} for instance). We neglect possible temperature dependence of CFPs\cite{Cadogan1988}.

$\hat{H}_{ext}=-\bold{H}_{ext}\cdot \bold{\hat{M}}_{4f}$ is the Zeeman interaction between the total 4$f$ moment operator $\bold{\hat{M}}_{4f}$ and the external magnetic field $\bold{H}_{ext}$.

\begin{figure}[h!]
\centering
\includegraphics[scale=0.21]{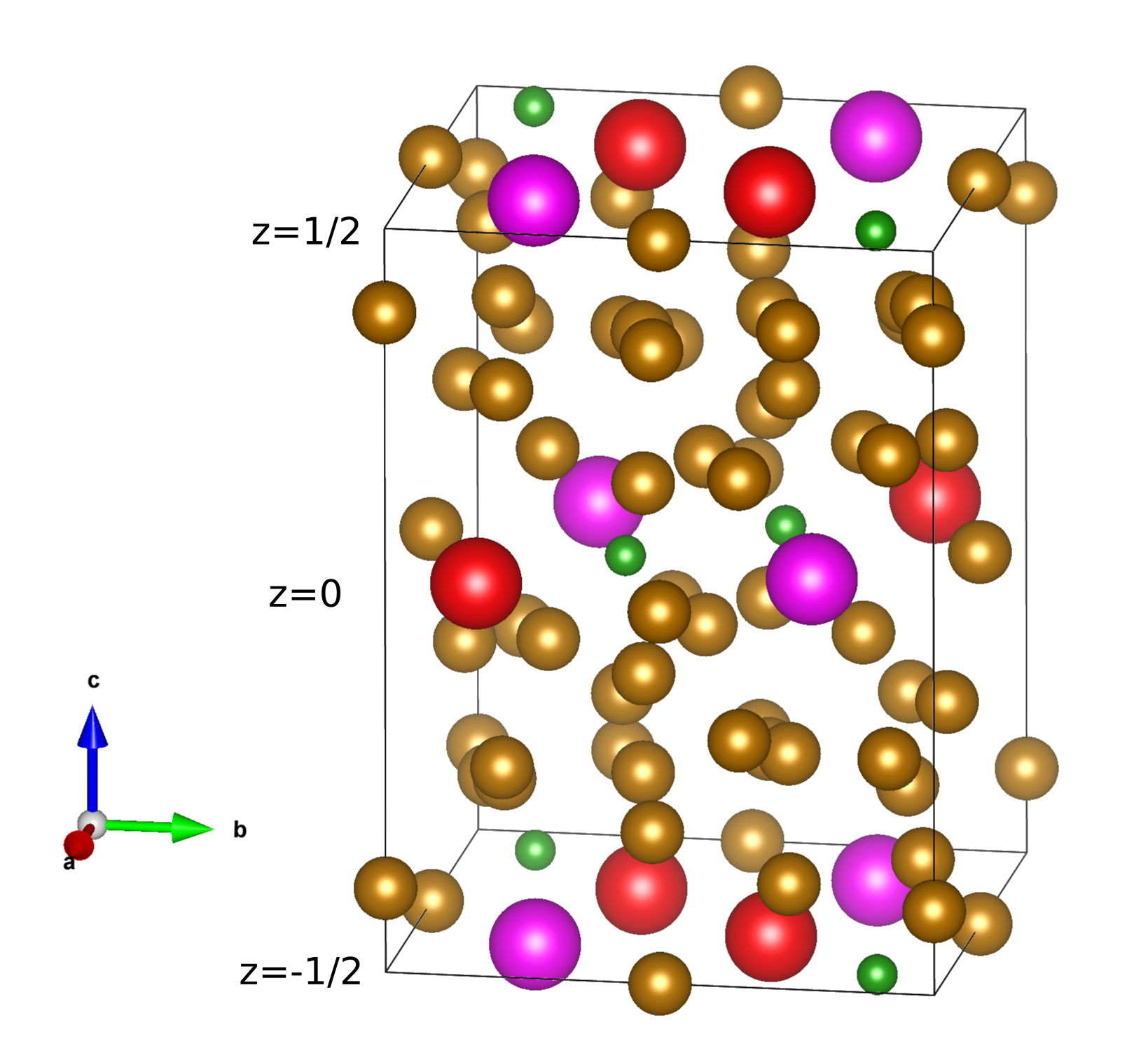}
\caption{2-14-1 crystal structure showing the different z planes and inequivalent R sites discussed in the main text: R $f$ site in purple, R $g$ site in red (Fe: large green, B: small green). These sites are also often labeled as 4$f$ and 4$g$ in the literature but we keep $f$ and $g$ in this work to avoid confusion with the notation for the R 4$f$ electrons. Plotted with VESTA\cite{Vesta}.}
\label{Struct}
\end{figure}

With the M sub-lattice magnetism treated by zero-temperature LSDA, the term $\hat{H}_{ex}$ is obtained for $T=$0 (and the external field $\bold{H}_{ext}$ is zero). The CFPs, spin-orbit coupling and exchange field are extracted this way for the two inequivalent R sites, labeled $f$ and $g$, as well as for the two $z=0$ and $z=1/2$ planes\cite{Herbst1991} (cf. Fig. \ref{Struct}). To remove the unphysical 4$f$ contribution to the CF splitting and exchange field, the DFT+HI scheme described above also employs a self-interaction correction; the computed CFPs are furthermore weakly dependent on the values of $U$ and $J_H$\cite{Delange2017}.

\subsection{Two sub-lattice model of 3$d$-4$f$ intermetallics}\label{subsec:twosublat}

The results of the previous sub-section (i.e. one-electron 4$f$ Hamiltonian defined in (\ref{H4f}), with its parameters calculated \textit{ab initio} for each R site in the unit cell) are then employed within the two sub-lattice picture of rare-earth transition-metal intermetallics. The latter decomposes the total free energy $F$ of a (R$_{1-x}$R'$_{x}$)$_{2}$Fe$_{14}$B system as:

\begin{align}
    F(T)& =\sum_{i}w_{i}F_{4f}^{(i)}(T)+F_{3d}(T) \label{F}\\
    F_{3d}(T)&=-\boldsymbol{M}_{3d}(T) \cdot \boldsymbol{H}_{ext}+K_{1}^{3d}(T)\sin^{2}\theta_{3d} \label{F3d}\\
    F_{4f}^{(i)}(T)&=-T\log\Bigg(\Tr{\exp{-\beta(\hat{H}_{1el}^{(i)}+\hat{H}_U)}}\Bigg) \label{F4f}
\end{align}
where $w_{i}$ is the occupancy of magnetic (Nd or Dy) ion $i=$ (R, site, plane), $K_{1}^{3d}(T)$ is the first anisotropy constant of the M sub-lattice, $\theta_{3d}$ is the angle between the M moment $\boldsymbol{M}_{3d}(T)$ and the [001] axis,  $\beta=1/(k_{B}T)$. 
The calculation of the 4$f$ free energy (\ref{F4f}) is performed by taking into account states up to the first excited multiplet, whose importance has been pointed out in previous works\cite{Pourovskii2020,Yamada1988}. Magnetization curves can then be obtained by minimizing (\ref{F}) with respect to $\theta_{3d}$. In the following subs-sections \ref{subsec:M3d} and \ref{subsec:M3d_vs_T}, we explain how we set the parameters describing the 3$d$ sub-lattice in Eqs. (\ref{F})-(\ref{F4f}), namely $M_{3d}(T)$, $B_{ex}(T)$ and $K_{1}^{3d}(T)$.

\subsection{3$d$ sub-lattice at zero  temperature}\label{subsec:M3d}

We assume the zero temperature $K_{1}^{3d}$ to be independent on the R ions and evaluate it in La$_{2}$Fe$_{14}$B by performing two separate LSDA+U calculations with the M moment aligned along [001] and [100], and $U-J=1.1$~eV for the M 3$d$ shells. We then employ the method of Ref. \onlinecite{Liu2020} computing 
$K_1^{3d}$ from the change of M sub-lattice spin-orbital energy upon its moment rotation: 
\begin{equation}
K_1^{3d}=\sum_i \Delta E^{SO}_i/2,
\end{equation}
where the sum runs over all M ions, $\Delta E^{SO}=E^{SO}_i(\boldsymbol{M}_{3d}||100)-E^{SO}_i(\boldsymbol{M}_{3d}||001)$. The spin-orbit energy $E^{SO}_i$ for site $i$  is calculated as $\mathrm{Tr}\left[\rho^{3d}_i \hat{H}^{3d}_{SO}\right]$, where $\rho^{3d}_i$ is the on-site 3$d$ density matrix for a given $\boldsymbol{M}_{3d}$ direction and the spin-orbit Hamiltonian $\hat{H}^{3d}_{SO}$ is of the same form as for the R shell but with the LSDA+U-estimated value $\lambda_{Fe}=60$~meV. This approach yields the value  $K_{1}^{3d}=0.4$~MJ.m\textsuperscript{-3} for La$_{2}$Fe$_{14}$B, in reasonable agreement with  experimentally measured value\cite{Bolzoni1987} $K_{1}^{3d}=0.7$~MJ.m\textsuperscript{-3}.

At 0 K, both $B_{ex}$ and the 3$d$ spin moment are extracted from the material specific DFT+HI scheme described in \ref{subsec:estruct}. Namely, $B_{ex}$ is extracted from the converged one-electron 4$f$ Hamiltonian as described in \ref{subsec:SIM}. The 3$d$ spin moment $S_{3d}$ is calculated  simultaneously. Notice, that since the R spin-polarization is suppressed in the calculated DFT+HI charge density, as described in Ref.~\onlinecite{Delange2017}, the resulting DFT+HI total spin moment corresponds to $S_{3d}$. It is found to be essentially the same ($30$~$\mu_{B}$/f.u.) in Nd$_{2}$Fe$_{14}$B and Dy$_{2}$Fe$_{14}$B. Since $B_{ex}$ originates from the same 3$d$ spin polarization in the same calculation, this treatment ensures consistent values of these parameters.

We expect the 3$d$ zero-temperature orbital moment $L_{3d}$ to be essentially independent on the R ions in the system. We therefore add to $S_{3d}$ the value of $L_{3d}$ computed by LSDA+U in La$_{2}$Fe$_{14}$B, yielding $0.06$~$\mu_{B}$/atom (in reasonable agreement with the experimental value of $\sim0.08$ $\mu_{B}$/atom\cite{Garcia2000}) and a total moment per unit cell of $M_{3d}=L_{3d}+S_{3d}=30.8$ $\mu_{B}$ at zero temperature.

\subsection{Temperature scaling of the 3$d$ sub-lattice}\label{subsec:M3d_vs_T}

Consistently with the definition of the exchange field, $\bold{M}_{3d}(T)$ and $\bold{B}_{ex}(T)$ are assumed to be anti-aligned (as is the case in our calculations for $T=$0) and, neglecting the temperature variation of the exchange coupling constant, to follow the same temperature scaling, for which we use Kuz'min's semi-empirical one\cite{Kuzmin2005}:
\begin{align}
    M_{3d}(T)& =M_{3d}\alpha(T)\\
    B_{ex}(T)& =B_{ex}\alpha(T)\\
    \alpha(T) & = \Bigg[1-s\left(\frac{T}{T_{c}}\right)^{\frac{3}{2}}-(1-s)\left(\frac{T}{T_{c}}\right)^{p} \Bigg]^{\frac{1}{3}}
\end{align}
where $T_{c}$ is the Curie temperature taken from experiment\cite{Herbst1991} for pure compounds and computed as the weighted sum of parent pure compounds for mixed systems (cf. Table \ref{Params}). We use $s=0.6$ and $p=5/2$ in accordance with Ref. \onlinecite{Kuzmin2010}.

$K_{1}^{3d}(T)$ is assumed to follow the temperature scaling of Zener\cite{Zener1954} (although it is a rather simplistic approximation in 2-14-1 systems, as shown in Refs. \onlinecite{Andreev1986,Bolzoni1987,Hirosawa1986}):
\begin{equation*}
    K_{1}^{3d}(T)=K_{1}^{3d}\Big(M_{3d}(T)/M_{3d}\Big)^3
\end{equation*}
which gives $K_{1}^{3d}(T)=K_{1}^{3d}\alpha(T)^{3}$.

\subsection{Treatment of Ce-based compounds}\label{subsec:Ce}

A special treatment is necessary for Ce-based compounds which we explain here. Indeed, according to the  measurements of Ref. \onlinecite{Colin2016}, Ce is found in an intermediate valence state dominated by Ce$^{4+}$ in "2-14-1" intermetallics, the description of which would require the use of more sophisticated and computationally heavy many-body approaches such as Quantum Monte-Carlo\cite{Galler2021}. Moreover, in 2-14-1 systems with localized rare-earth, those localized R ions (Nd, Dy etc.) provide a dominant contribution to the magnetic anisotropy. The contribution of Ce, whether with an itinerant or localized 4$f$ shell, is expected to be relatively small, in particular at  room temperature and above (as shown in Ref.~\onlinecite{Galler2021}). Therefore, in mixed compounds (Nd$_{1-x}$Ce$_{x}$)$_{2}$Fe$_{14}$B:\\
\begin{itemize}
    \item Ce is treated within LSDA for CFP calculations of Nd by the DFT+HI scheme described above;
    \item Ce contributions to the magnetic moment and anisotropy in the two sub-lattice model are described by a mere renormalization of the same $3d$ quantities at zero temperature. Specifically, as the total experimental moment per unit cell in Ce$_{2}$Fe$_{14}$B is $29.4$~$\mu_{B}$ at 4.2 K\cite{Herbst1991}, we adapt $M_{3d}=30.8(1-x)+29.4x$~$\mu_{B}$ per unit cell. Furthermore, as we measured a larger zero temperature $K_{1}=1.6$~MJ.m\textsuperscript{-3} in Ce$_{2}$Fe$_{14}$B (in agreement with previous measurements\cite{Bolzoni1987,Hirosawa1986}) compared to La$_{2}$Fe$_{14}$B, we adapt $K_{1}^{3d}=0.4(1-x)+1.6x$~MJ.m\textsuperscript{-3} (cf. Table~\ref{Params}).
\end{itemize}

\begin{table}[!h]
\begin{center}
\begin{tabular}{l| c c}
  \hline
  \hline
     & $T_{c}$  & $K_{1}^{3d}$ \\
  \hline
    Nd$_{2}$Fe$_{14}$B  & 585  & 0.4\\
      \hline
         Ce$_{2}$Fe$_{14}$B  & 424  & 1.6 \\
      \hline
        (Ce$_{0.63}$Nd$_{0.37}$)$_{2}$Fe$_{14}$B  & 484 & 1.16 \\
         \hline
        (Ce$_{0.36}$Nd$_{0.64}$)$_{2}$Fe$_{14}$B  & 527 & 0.83\\
  \hline
        Dy$_{2}$Fe$_{14}$B & 598  & 0.4 \\
    \hline
            (Dy$_{0.36}$Nd$_{0.64}$)$_{2}$Fe$_{14}$B & 590  & 0.4 \\
  \hline
  \hline
\end{tabular}
 \caption{\label{Params} Values of the Curie temperature $T_{c}$ (in K) and of the 3d first anisotropy constant at 0 K $K_{1}^{3d}$ (in MJ.m\textsuperscript{-3}) involved in the two sub-lattice model for different compounds.}
 \end{center}
\end{table}

\section{Experimental methods}
(Nd$_{1-x}$Ce$_{x}$)$_{2}$Fe$_{14}$B (x=0, 0.63, 1) single crystals were grown using the “reactive flux” method as described in our previous study~\cite{Gomez2021}. 

The chemical composition was determined using
Energy Dispersive X-ray (EDX) spectroscopy equipped in a Philips XL30
FEG Scanning Electron Microscopy (SEM). The single crystallinity and
crystallographic orientation of the single crystals were verified using a
back-scattering Laue camera. The magnetization measurement was performed using a Physical properties Measurement System (PPMS
14, Quantum Design) with vibrating Sample Magnetometer (VSM) up to 14 T at temperatures ranging from 10 to 300 K. For magnetization measurements at
fields up to 50 T, a pulsed-field magnetometer built at the High-field
Laboratory in Dresden-Rossendorf (HZDR) was used, as described in detail in Ref. \onlinecite{skourski2011}.
The MAE was extracted from the experimental results by calculating the integral of the M(H) curves for [100] and [001] axis (hard and easy axis at RT, respectively), and we subtracted the integral of the [001] (which is taken as reference, also below spin reorientation) from the one of [100].

\section{Results}

\subsection{Crystal-field parameters and exchange field}
\begin{figure}[h!]
\centering
\includegraphics[scale=0.3]{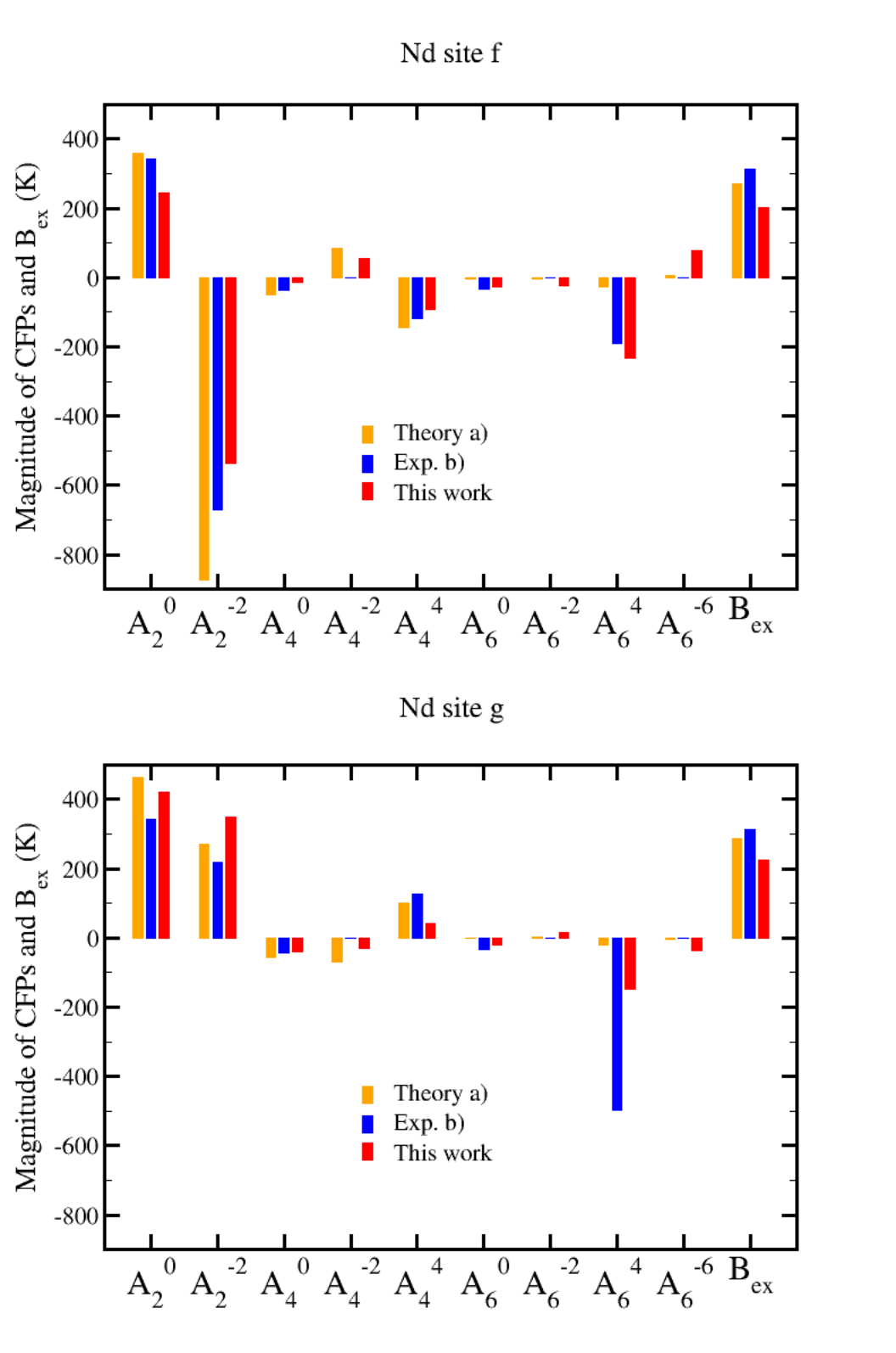}
\caption{Nd CFPs and $B_{ex}$ (in K) for both inequivalent sites in the 2-14-1 system. The values are for the $z=0$ plane of the unit cell (the signs of $A_{2}^{-2}\expval{r^{2}}$, $A_{4}^{-2}\expval{r^{4}}$, $A_{6}^{-2}\expval{r^{6}}$ and $A_{6}^{-6}\expval{r^{6}}$ change for the middle plane). a) Ref. \onlinecite{Sato2021}. b) Ref. \onlinecite{Cadogan1988} (Ref. \onlinecite{Li1988} for $B_{ex}$).}
\label{CFPs}
\end{figure}

Nd CFPs and $B_{ex}$ computed within our approach (cf. \ref{subsec:estruct} and \ref{subsec:SIM}) in Nd$_{2}$Fe$_{14}$B, together with experimental\cite{Cadogan1988} and previous \textit{ab initio}\cite{Sato2021} values, are summarized on Figure \ref{CFPs} for $f$ and $g$ inequivalent sites in the unit cell. Our theoretical parameters have the same sign and order of magnitude than the  ones extracted by Ref.~\onlinecite{Cadogan1988} from experimental magnetization curves (only a subset of CFPs was assumed to be non-zero in their fitting). 
The only significant discrepancy with respect to Ref.~\onlinecite{Cadogan1988} is the underestimation of $A_{6}^{4}\langle r^{6}\rangle$ on the $g$ site. Overall, the results of our \textit{ab initio} approach are comparable with those of Ref. \onlinecite{Sato2021}.

The precise values are summarized in Table \ref{CFPstable} in the Appendix, which also lists computed CFPs for Dy as well as our results 
for various mixed systems with Ce or La (both treated within LSDA) occupying one of the two sites. These calculations with partial substitution give, for Nd and Dy, essentially the same CFPs and $B_{ex}$ as in the corresponding pure compounds. This shows that the CFPs on one R site are insensitive to substitution of the R element  at the other site, therefore justifying the use of equation (\ref{F}).

Furthermore, in all cases, the $g$ site exhibits higher or equal $A_{2}^{0}\langle r^{2}\rangle$ and $B_{ex}$ compared to the $f$ site. This suggests that the $g$ site exhibits a higher SIA than the $f$ one (in agreement with the results of Refs.~\onlinecite{Chouhan2018,Sato2021}), at least at high temperature, where higher order CFPs are essentially negligible. This idea will be explored in the last two subsections.

\subsection{Magnetic properties of pure Nd$_{2}$Fe$_{14}$B}

We employ the computed CFPs shown in Figure \ref{CFPs} and parameters summarized in Table \ref{Params} within the two sub-lattice picture (cf. \ref{subsec:twosublat}, \ref{subsec:M3d} and \ref{subsec:M3d_vs_T}) to compute 
magnetic properties of Nd$_{2}$Fe$_{14}$B and  compare our predictions with experimental data.

\begin{figure}[h!]
\centering
\includegraphics[scale=0.3]{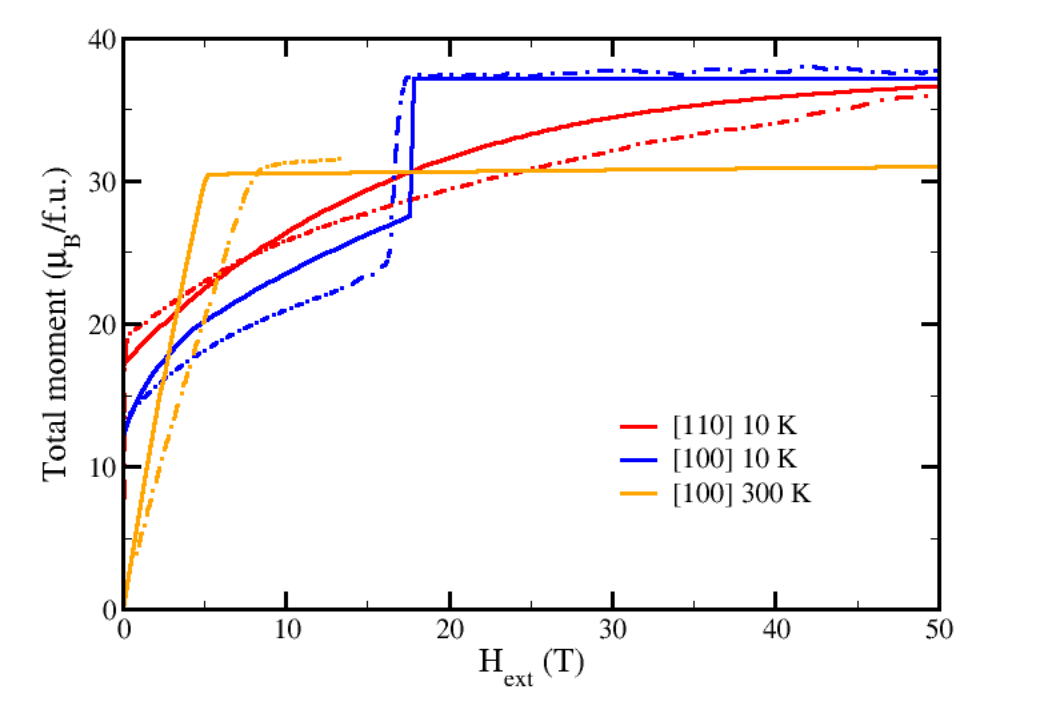}
\caption{Nd$_{2}$Fe$_{14}$B magnetization curves, along [100] and [110], at T=$10$ and $300$ K. Dotted lines: experiment\cite{Gomez2021}. Solid lines: theory.}
\label{MvH}
\end{figure}

Fig.~\ref{MvH} displays the computed Nd$_{2}$Fe$_{14}$B magnetization curves along the [100] and [110]  directions together with the experimental ones\cite{Gomez2021} at T=$10$ K and $300$ K. The experimental curves are well reproduced, including some subtle features such as the First-Order Magnetization Process (FOMP) along [100] at $T=10$ K (at $H_{ext}=17$~T) as well as the saturation ($\sim 37$ $\mu_{B}$/f.u. along [100] at $10$ K) and spontaneous magnetizations ($\sim 13$ $\mu_{B}$/f.u. and $\sim 17$ $\mu_{B}$/f.u. at $10$ K along [100] and [110] respectively).

\begin{figure}[h!]
\centering
\includegraphics[scale=0.3]{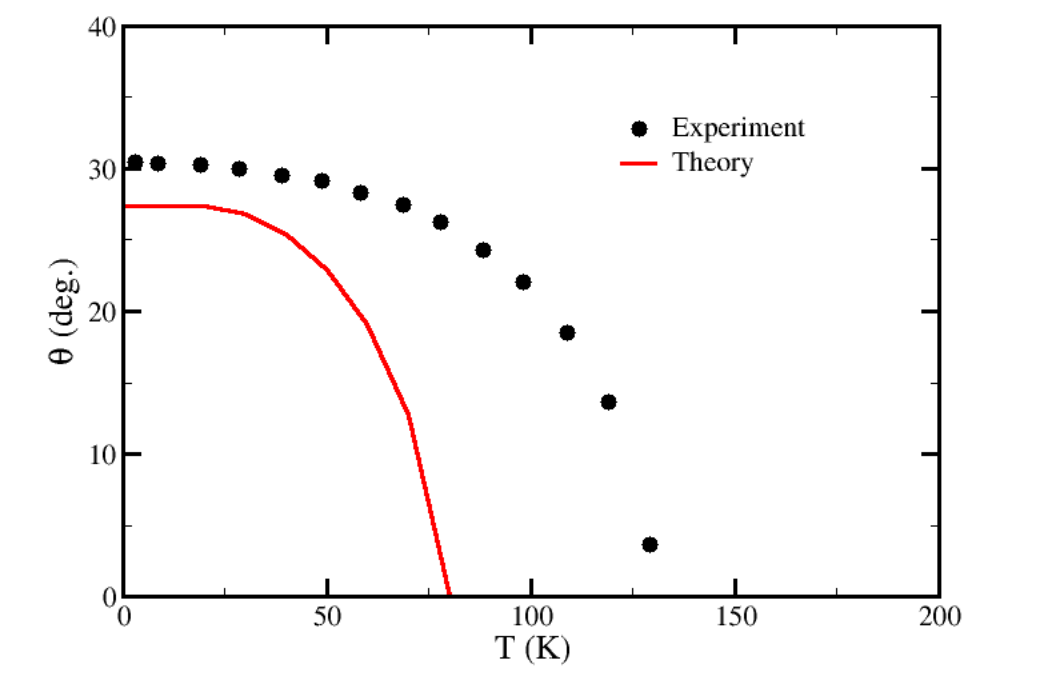}
\caption{Temperature evolution of the angle $\theta$ between the total moment and the [001] axis (in the (110) plane) in Nd$_{2}$Fe$_{14}$B. Experiment is from Ref. \onlinecite{Cadogan1988}.}
\label{theta}
\end{figure}

Fig.~\ref{theta} shows the evolution of the angle $\theta$ between the total spontaneous magnetization and the [001] axis (in the (110) plane) in Nd$_{2}$Fe$_{14}$B. Our calculations reproduce the spin reorientation transition (SRT) occurring at low temperature due to the competition. 
The low temperature angle is also reproduced ($\sim30^{\circ}$), which is consistent with the excellent agreement of the low temperature magnetizations at zero field (Fig.~\ref{MvH}). The SRT temperature is however underestimated ($\sim 75$ K instead of $135$ K). There has been a lot of debate regarding the magnetic structure of the compound at $4$ K: some works\cite{Cadogan1988,Yamada1988,Nowik1990} predicted a very small canting angle between Nd moments and the total one ($\leq 7^{\circ}$ ), while others suggested a large one\cite{Onodera1987,Bartolome2000,Garcia2000_bis,Wolfers2001}. Our calculations support the small canting angle picture, with a maximum angle of $5^{\circ}$  between a Nd moment and the total one.

\begin{figure}[h!]
\centering
\includegraphics[scale=0.3]{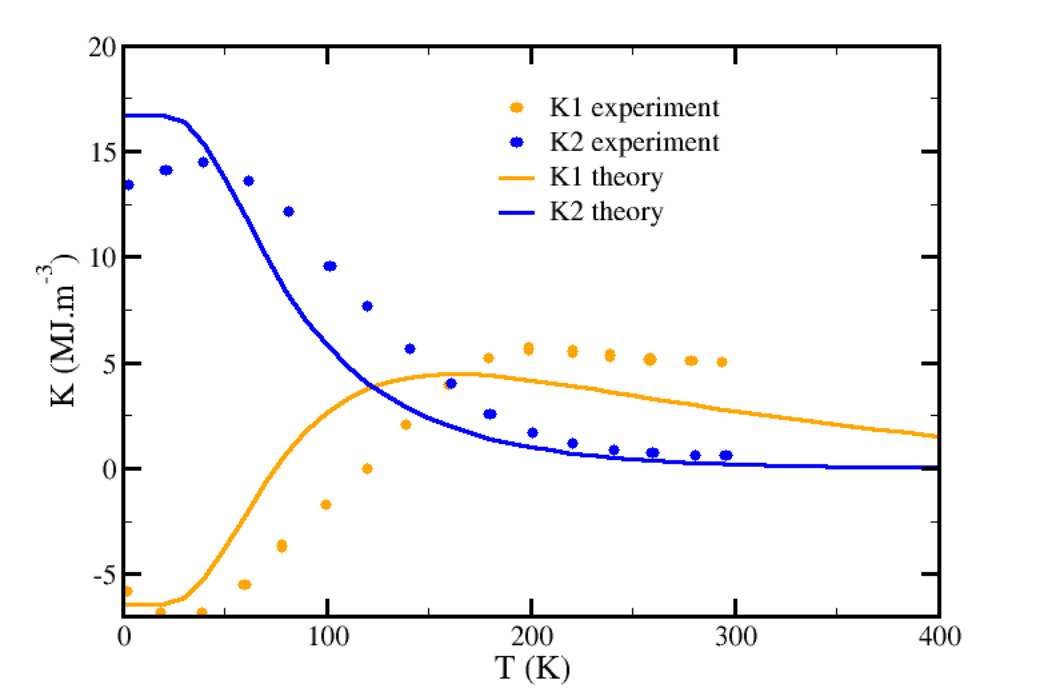}
\caption{Evolution of Nd$_{2}$Fe$_{14}$B anistropy constants $K_{1}$ and $K_{2}$ with temperature, extracted by the Sucksmith-Thompson method\cite{Sucksmith1954}. The experimental data was taken from Ref. \onlinecite{Andreev1986}.}
\label{Ks}
\end{figure}

Fig.~\ref{Ks} displays the evolution of Nd$_{2}$Fe$_{14}$B anisotropy constants $K_{1}$ and $K_{2}$ extracted with the Sucksmith-Thompson (ST) method\cite{Sucksmith1954}. Although it assumes perfectly aligned Fe and Nd moments, we employ the ST method in order to have a consistent comparison with experimental anisotropy constants that were also extracted with it. The agreement is fairly good. It is also consistent with Fig.~ \ref{theta}: at high temperatures, both anisotropies constant are positive and the phase is therefore uniaxial. At low temperature, the competition between negative $K_{1}$ and positive $K_{2}$ results in the conical phase. The temperature at which $K_{1}$ changes signs is underestimated, consistently with the underestimation of the spin reorientation transition temperature (Fig. \ref{theta}).

Our method therefore proves to be a reliable \textit{ab initio} approach to the magnetic properties of the complicated Nd$_{2}$Fe$_{14}$B system: it provides site-specific Nd CFPs as well as exchange field and explains different subtle experimental features (SRT, FOMP, etc.), with the crystal structure, temperature dependence of the 3d magnetic moment and Wannier projection window as the only parameters.

\subsection{Ce substitution}

\begin{figure}[h!]
\centering
\includegraphics[scale=0.3]{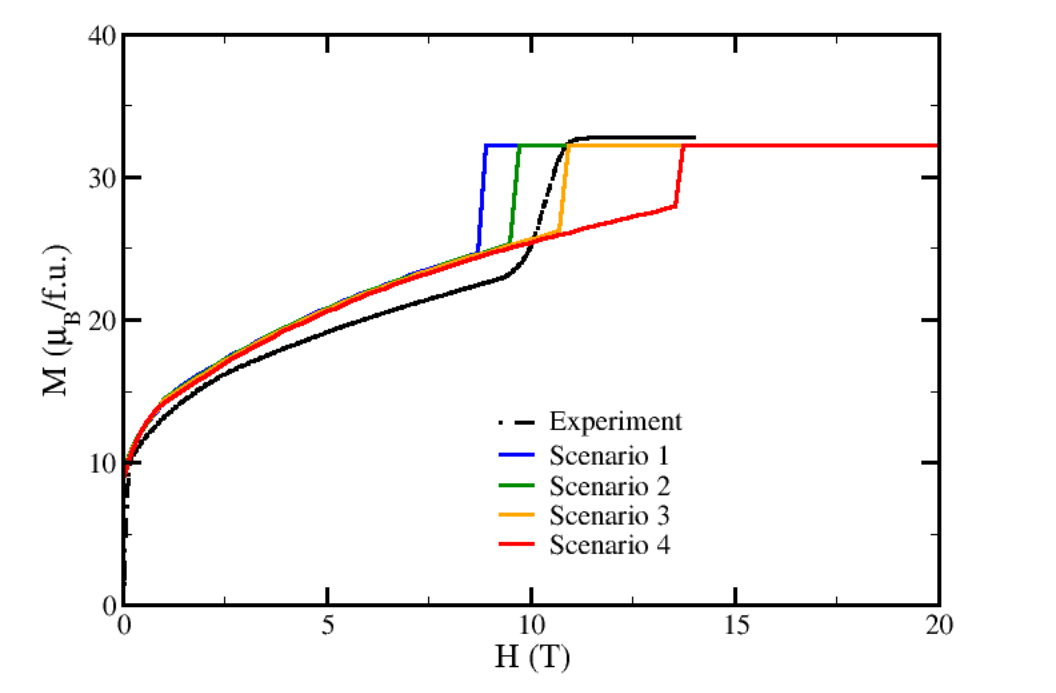}
\caption{Magnetization curve of (Nd$_{0.37}$Ce$_{0.63}$)$_{2}$Fe$_{14}$B along [100] at 10 K, according to experiment and computed for the different occupancy scenarios 1-4 (cf. Table \ref{Occ}).}
\label{Ce63}
\end{figure}

Within the two sub-lattice model (cf. \ref{subsec:twosublat}, \ref{subsec:M3d}, \ref{subsec:M3d_vs_T}, \ref{subsec:Ce}), we now turn to partial substitution of Nd by lighter, more abundant and cheaper Ce whose preferred site is still controversial. Indeed, Ref. \onlinecite{Colin2016} showed that Ce favors the smaller $f$ site in their compounds (consistently with a valence dominated by Ce$^{4+}$). However, Ref. \onlinecite{Lin2020} suggested the opposite. To investigate this issue, we compared the experimental FOMP at 10 K along [100] in (Nd$_{0.37}$Ce$_{0.63}$)$_{2}$Fe$_{14}$B with theoretical calculations for different occupancy scenarios (cf. Table \ref{Occ}): Ce with a slight $g$ preference as measured by Ref. \onlinecite{Lin2020} (scenario 1), no site preference between Ce and Nd (scenario 2), Ce with a slight $f$ preference as measured by Ref. \onlinecite{Colin2016} (scenario 3) and Ce fully occupying the $f$ site (scenario 4). 
As illustrated on Figure \ref{Ce63}, gauged by the FOMP field, scenario 3 gives the best agreement with experiment. However, one may notice that the relative error in the determination of the FOMP field for the pure compound (about 2 Tesla) suggests an accuracy of about 1 Tesla for the mixed case with the error scaled correspondingly to the lower Nd concentration. Within this uncertainty, we cannot discriminate between the scenarios 2 and 3; however, Nd preference for $f$ site (scenario 1) and  purely $g$ Nd  occupancy (scenario 4) are unlikely. Therefore, the experimental situation likely corresponds to either no preference or a slight Ce preference for the $f$ site.
This implies that there is potentially room for site occupancy optimization which will be studied below.

\begin{table}[!h]
\begin{center}
\begin{tabular}{l c | c c c c}
  \hline
  \hline
     Compound & Scenario  &  Nd $f$ & Nd $g$ & R $f$ & R g\\
       \hline
    (Nd$_{0.37}$Ce$_{0.63}$)$_{2}$Fe$_{14}$B   & 1. Ref. \onlinecite{Lin2020}  & 0.45& 0.29 & 0.55 & 0.71\\
     & 2. No pref.  & 0.37 & 0.37 &0.63 & 0.63\\
     & 3. Ref. \onlinecite{Colin2016}  &  0.25 & 0.49 & 0.75 & 0.51\\
     & 4. Ce on $f$  &  0.0 & 0.74 &1.0 & 0.26\\
        \hline
    (Nd$_{0.64}$Ce$_{0.36}$)$_{2}$Fe$_{14}$B  & 5. Ref. \onlinecite{Colin2016}  &  0.56 & 0.72 & 0.44 & 0.28\\
     & 6. Ce on $f$  &  0.28 & 1.0 & 0.72 & 0.0\\
  \hline
      (Nd$_{0.64}$Dy$_{0.36}$)$_{2}$Fe$_{14}$B  & 7. Ref. \onlinecite{Saito2017} &  0.51 & 0.77 & 0.49 & 0.23\\
       & 8. Dy on $g$ & 1.0 & 0.28 & 0.0 & 0.72 \\
  \hline
  \hline
\end{tabular}
 \caption{\label{Occ}Site-detailed stoichiometry of (Nd$_{1-x}$R$_{x}$)$_{2}$Fe$_{14}$B compounds (R=Ce, Dy), depending on various occupancy scenarios. Numbers labeling the scenarios are used for clear reference in the main text and figures. No pref. = no site preference between Nd and R. R on i = ion R occupies first the site i. Ref. \onlinecite{Lin2020} = Ce has a slight preference for the $g$ site. Ref. \onlinecite{Colin2016} = Ce has a slight preference for the $f$ site. Ref. \onlinecite{Saito2017} = Dy has a slight preference for the $f$ site. The actual occupation numbers of scenarios 1, 3, 5 and 7 were extracted by interpolation of the measured data displayed in the respective references.}
 \end{center}
\end{table}

\begin{figure}[h!]
\centering
\includegraphics[scale=0.3]{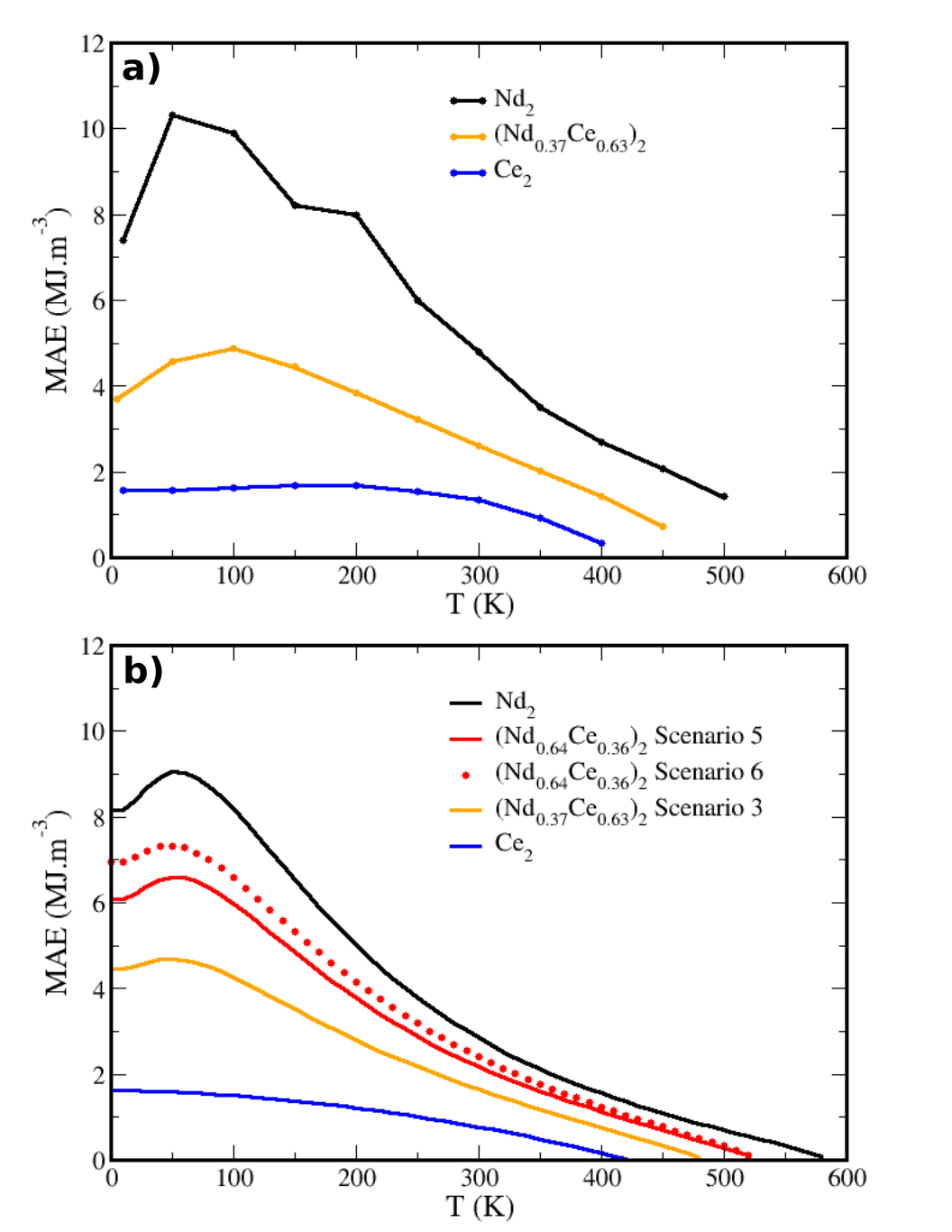}
\caption{Magnetocrystalline anisotropy as a function of temperature of (Nd$_{1-x}$Ce$_{x}$)$_{2}$Fe$_{14}$B compounds. a): experiment. b): theory according to occupancy scenarios of Table \ref{Occ}. In the case of Ce$_{2}$Fe$_{14}$B, the theoretical MAE amounts to $K_{1}^{3d}(T)$ as explained in the Methods section.}
\label{MAECe}
\end{figure}

We then computed and measured the magnetocrystalline anisotropy MAE $=F_{[100]}-F_{[001]}$ for different Ce concentration: Nd$_{2}$Fe$_{14}$B, (Nd$_{0.64}$Ce$_{0.36}$)$_{2}$Fe$_{14}$B, (Nd$_{0.37}$Ce$_{0.63}$)$_{2}$Fe$_{14}$B and Ce$_{2}$Fe$_{14}$B. For the theoretical estimation, we directly evaluated the MAE from Equation (\ref{F}) using the experimental occupancy of Ref. \onlinecite{Colin2016} for the mixed systems (scenarios 3 and 5 in Table \ref{Occ}). The results are displayed on Figure \ref{MAECe}. The MAE decreases, as expected, as a function of the Ce concentration which contributes weakly to the MAE (only through the renormalization of $K_{1}^{3d}$ in this work) compared to Nd. The overall agreement between theory and experiment is fairly good which shows the ability of our approach to treat complex substituted systems.

To investigate the importance of site occupancy, we also computed the MAE with Ce occupying only the $f$ site, for the (Nd$_{0.64}$Ce$_{0.36}$)$_{2}$Fe$_{14}$ stoichiometry (scenario 6 in Table \ref{Occ}); the result is displayed on Figure \ref{MAECe}. Compared to the experimental occupancy of Ref. \onlinecite{Colin2016}, we computed an increase of anisotropy over the whole range of temperature (+9\% at 300 K for instance, from $2.2$ to $2.4$ MJ.m$^{-3}$). This effect arises from the larger Nd $g$ SIA due to larger $A_{2}^{0}\langle r^{2}\rangle$ and $B_{ex}$ (as discussed above, cf. Figure \ref{CFPs}). Therefore, by  further enhancing $f$ site Ce occupancy (and, correspondingly, Nd $g$ occupancy)
one should be able to increase
 the MAE of (Nd,Ce)$_2$Fe$_{14}$B  substituted compounds. It also means that when it comes to non-magnetic Nd substitutions, it is, in principle, preferable to use elements with smaller ionic radii in order to keep Nd at the $g$ site. 

\subsection{Dy substitution}

\begin{figure}[h!]
\centering
\includegraphics[scale=0.3]{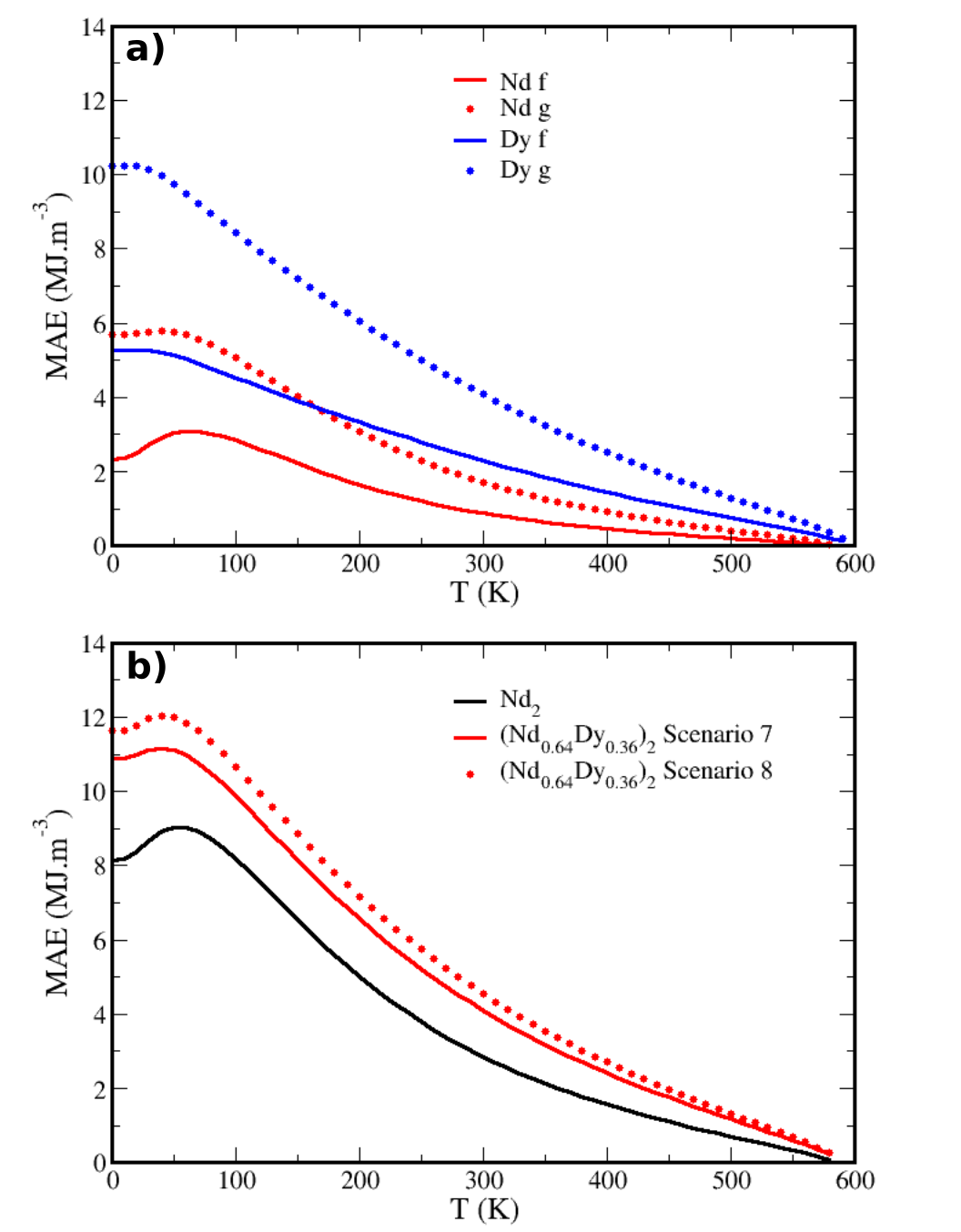}
\caption{Evolution of theoretical MAE with temperature. a): per (R, site). b): Nd$_{2}$Fe$_{14}$B and (Nd$_{0.64}$Dy$_{0.36}$)$_{2}$Fe$_{14}$B according to occupancy scenarios of Table \ref{Occ}.}
\label{MAEDy}
\end{figure}

Within the two sub-lattice model (cf. \ref{subsec:twosublat}, \ref{subsec:M3d}, \ref{subsec:M3d_vs_T}), we now turn to Dy substitution, which is routinely used in the industry to enhance anisotropy in Nd$_{2}$Fe$_{14}$B. According to Ref. \onlinecite{Saito2017}, Dy has a preference for the $f$ site, which is consistent with its smaller atomic radius compared to Nd (lanthanide contraction). As in the case of Ce, the experimental occupancy is still close to equal occupation between sites, but the situation is reversed: here, heavy R occupying the $f$ site is expected to be detrimental to the anisotropy. Indeed, as shown on Figure \ref{MAEDy}a), Dy has a larger SIA and also features a larger contribution arising from the $g$ site. It results in MAE$_{Dy}^{g}$-MAE$_{Dy}^{f}$ $>$  MAE$_{Nd}^{g}$-MAE$_{Nd}^{f}$ which means that, for a given Dy content, forcing Dy to occupy the $g$ site is predicted to enhance the anisotropy. This is illustrated on Figure \ref{MAEDy}b): theoretical MAE of (Nd$_{0.64}$Dy$_{0.36}$)$_{2}$Fe$_{14}$B with the experimental occupancy from Ref. \onlinecite{Saito2017} (scenario 7 in Table \ref{Occ}) is larger than in pure Nd$_{2}$Fe$_{14}$B but can be further increased by assuming Dy to occupy the $g$ site (scenario 8 in Table \ref{Occ}). We computed for instance an increase of 10\% at 300 K (from 4.1 to 4.5 MJ.m$^{-3}$). Therefore, were we able to force Dy on the $g$ site in Nd-Dy compounds, we could increase, though moderately, the anisotropy per Dy content. The same kind of phenomenon is expected to occur with Tb substitution.

\section{Conclusion}

In this work, we compute from a first-principles DFT+DMFT approach site specific crystal-field and exchange field parameters of Nd in the theoretically challenging Nd$_{2}$Fe$_{14}$B intermetallic, which is the most widely used high-performance hard magnet in the industry. The resulting values are in excellent agreement with previously computed and experimental ones. Moreover, we show that these parameters are essentially insensitive to  substitutions on the other R site in the 2-14-1 structure, an assumption usually made in the literature but never proven. 
We also
show that with these parameters one may construct an almost fully \textit{ab initio} two sub-lattice model that  reproduces various measured magnetic properties of Nd$_{2}$Fe$_{14}$B. 

Next, we study  industrially motivated partial substitution of Nd by Ce or Dy, focusing on the substitution site preference and its impact on magnetic properties.
In Ce substituted compounds -- often considered in the quest to reduce the scarce heavy rare-earth content in hard magnets -- comparison of experimental data\cite{Colin2016,Lin2020} with our theoretical calculations supports a slight Ce preference for the $f$ site. We carry out  experimental measurements of MAE in Ce substituted single crystals, which are found to be in good agreement with our theoretical predictions. Our calculations also predict that enhancing Nd preferential g-site occupancy leads to a higher SIA. 
This theoretical observation implies that one might be able to enhance the magnetic anisotropy in (Nd,Ce)$_2$Fe$_{14}$B compounds by engineering higher  Ce f-site occupancy.

In Dy substituted compounds, routinely used in the industry, we also show that an increase -- though moderate -- of the magnetic anisotropy is theoretically expected when Dy 
occupies the $g$ site. This could perhaps be achieved by adding a third R element with an even smaller radius (therefore occupying preferably the $f$ site), such as Ce$^{4+}$: while this kind of doping would lower the anisotropy of the compound, it could increase the anisotropy per Dy atom by the mechanism aforementioned as well as reduce the total economic cost (as Ce is cheaper than Dy), resulting in an overall better anisotropy/price ratio.

Overall, our \textit{ab initio} approach 
is shown to be a reliable tool to predict and analyze 
intrinsic 
properties of complex, substituted, hard magnetic materials. It may provide useful insight into site occupancy optimization and its impact on magnetic properties.

\section{Acknowledgements}
This work was supported by the future pioneering program "Development of magnetic material technology for high-efficiency motors" (MagHEM) commissioned by the New Energy and Industrial Technology Development Organization (NEDO). A. Aubert and K. Skokov acknowledge the financial support from the Deutsche Forschungsgemeinschaft - Project-ID 405553726 -TRR 270. L. Pourovskii and J. Boust are grateful to the CPHT computer support team. 

\section{Appendix}

Table \ref{CFPstable} summarizes CFPs and exchange fields computed in this work for various 2-14-1 systems.

\begin{table*}[!h]
\begin{center}
\begin{tabular}{l c|c c c c c c c c c c}
  \hline
    \hline
     & Site & $A_{2}^{0}\expval{r^{2}}$ & $A_{2}^{-2}\expval{r^{2}}$ & $A_{4}^{0}\expval{r^{4}}$ & $A_{4}^{-2}\expval{r^{4}}$ & $A_{4}^{4}\expval{r^{4}}$ & 
     $A_{6}^{0}\expval{r^{6}}$ &
     $A_{6}^{-2}\expval{r^{6}}$ &
     $A_{6}^{4}\expval{r^{6}}$ &
     $A_{6}^{-6}\expval{r^{6}}$ &
     $B_{ex}$\\

  \hline
    Nd$_{2}$Fe$_{14}$B   & $f$ & 246 & -537 & -14 & 54 & -91 & -27 & -24 & -232 & 78 & 203\\
             & $g$ & 420 & 349 & -41 & -29 & 42 & -19 & 15 & -149 & -37 & 222\\
              \hline
        NdLaFe$_{14}$B  & $f$ & 297 & -641 & -22 & 73 & -119 & -25 & -16 & -228 & 27 & 212\\

        CeNdFe$_{14}$B  & $g$ & 414 & 334 & -45 & -28 & 54 & -20 & 9 & -154 & -56 & 223\\
  \hline
     Dy$_{2}$Fe$_{14}$B  & $f$ & 113 & -345 & -4 & 29 & -19 & -12 & 27 &-93  &115  &209 \\
           & $g$ & 186 & 180 & -16 & -17 & 19 & -10 &-14  &-25  & 72 & 207\\
  \hline
     DyCeFe$_{14}$B  & $f$ & 115 & -381 & -6 & 32 & -21 & -14 & 32 & -95 & 123 & 189\\
     CeDyFe$_{14}$B      & $g$ & 190 & 129 & -18 & -7 & 8 & -11 & -21 & -15 &86  & 210\\
           
     \hline  
 
     \hline
\end{tabular}
\captionsetup{width=.9\textwidth}
 \caption{\label{CFPstable}Site specific theoretical CFPs and $B_{ex}$ (in K) for Nd and Dy computed in different compounds. The values are for the R ions in the $z=0$ plane of the unit cell (the signs of $A_{2}^{-2}\expval{r^{2}}$, $A_{4}^{-2}\expval{r^{4}}$, $A_{6}^{-2}\expval{r^{6}}$ and $A_{6}^{-6}\expval{r^{6}}$ change for the middle plane). In substituted compounds, Nd/Dy occupies the site indicated in the second column, which is also highlighted by the order of the ions in the compound formula (first ion on site $f$, second on site $g$).}
 \end{center}
\end{table*}

\end{document}